\documentclass{article}

\usepackage[dblblindworkshop, final]{neurips_2026}

\usepackage[utf8]{inputenc} % allow utf-8 input
\usepackage[T1]{fontenc}    % use 8-bit T1 fonts
\usepackage{hyperref}       % hyperlinks
\usepackage{url}            % simple URL typesetting
\usepackage{booktabs}       % professional-quality tables
\usepackage{amsfonts}       % blackboard math symbols
\usepackage{nicefrac}       % compact symbols for 1/2, etc.
\usepackage{microtype}      % microtypography       % colors
\usepackage{amsmath}
\usepackage{comment}
\usepackage{graphicx}
\usepackage{booktabs}
\usepackage{adjustbox}
\usepackage[table,dvipsnames]{xcolor}
\newcommand{\tablefit}[1]{%
  \begin{adjustbox}{max width=\linewidth,center}#1\end{adjustbox}%
}
\usepackage{array}
\usepackage{placeins}
\usepackage{float}
\usepackage{multirow}
\usepackage{float}
\usepackage{wrapfig}
\usepackage{tcolorbox}
\usepackage{listings}

\lstdefinestyle{promptstyle}{
  basicstyle=\ttfamily\scriptsize,
  breaklines=true,
  breakatwhitespace=true,
  columns=fullflexible,
  keepspaces=true,
  frame=single,
  rulecolor=\color{black!35},
  backgroundcolor=\color{gray!6},
  aboveskip=6pt,
  belowskip=8pt,
  framexleftmargin=4pt,
  framexrightmargin=4pt,
  showstringspaces=false,
  captionpos=b
}
\newcolumntype{L}[1]{>{\raggedright\arraybackslash}p{#1}}

\definecolor{svbestblue}{RGB}{226,235,249}

\title{Who Takes the Lead? Model Ability and Task-Conditioned Collaboration with Humans}

\author{
Weien Li\textsuperscript{1, 2}\thanks{
Corresponding authors:
\href{mailto:weien.li@mail.mcgill.ca}{weien.li@mail.mcgill.ca} and
\href{mailto:ye.yuan3@mail.mcgill.ca}{ye.yuan3@mail.mcgill.ca}.
}\And
Haochen Liu\textsuperscript{3}\And
Zeyu Li\textsuperscript{1}\And
Rui Song\textsuperscript{1, 4}\And
Zixuan Dong\textsuperscript{4}\And
Ye Yuan\textsuperscript{1, 5}\footnotemark[2]\AND
\textsuperscript{1} McGill University,
\textsuperscript{2} MIT - Massachusetts Institute of Technology,\AND
\textsuperscript{3} University of Cambridge,
\textsuperscript{4} University of Toronto,
\textsuperscript{5} Mila - Quebec AI Institute
}

\begin{document}

\setcounter{footnote}{1}
\maketitle

% https://neurips.cc/public/guides/PaperChecklist
%  Briefly mention Colpali Visual Retrieval Internal Representation and Late Interaction Limitations and how our idea is better. 
\begin{abstract}
As language-model agents grow more capable, they can carry more of a task without asking anyone for help. 
Falling reliance on a human partner is easy to read as a failure of collaboration when it may instead be a correct division of labor, and the rate of human involvement alone cannot tell the two apart. 
We study this systematically by placing an agent and a person under a shared objective while varying who holds the advantage: the human has a clear advantage, neither does, or the agent does. 
To obtain this control, we introduce a shared-control maze, in which an agent-controlled and a human-controlled avatar share the objective of reaching a common target in as few team turns as possible.
The agent allocates each team turn between acting itself and handing the move to the person, and initial placement sets the human's shortest-path advantage while the task is otherwise unchanged.
We evaluate GPT~5.6~Sol, Command~A+, Qwen~3.8~27B, and Command~R, first measuring solo capability with a single avatar and full-map observability, then running co-play with human participants across the three advantage conditions.
Across the four models, stronger solo performance is associated with selective, route-conditioned human involvement: the models that solve the maze reliably on their own act for themselves almost without exception when the human holds no advantage, and involve the human on roughly $80\%$ of initial decisions when the human route is shorter. 
Command R, the model with substantially weaker solo performance, shows no such selectivity.
%
% Our source code is available at 
% \href{https://anonymous.4open.science/r/Who-Takes-the-Lead-694F}{this anonymous repository}.
\end{abstract}
\section{Introduction}
\label{sec:intro}

\begin{figure}[t]
    \centering
    \vspace{-10mm}
    \includegraphics[width=\linewidth]{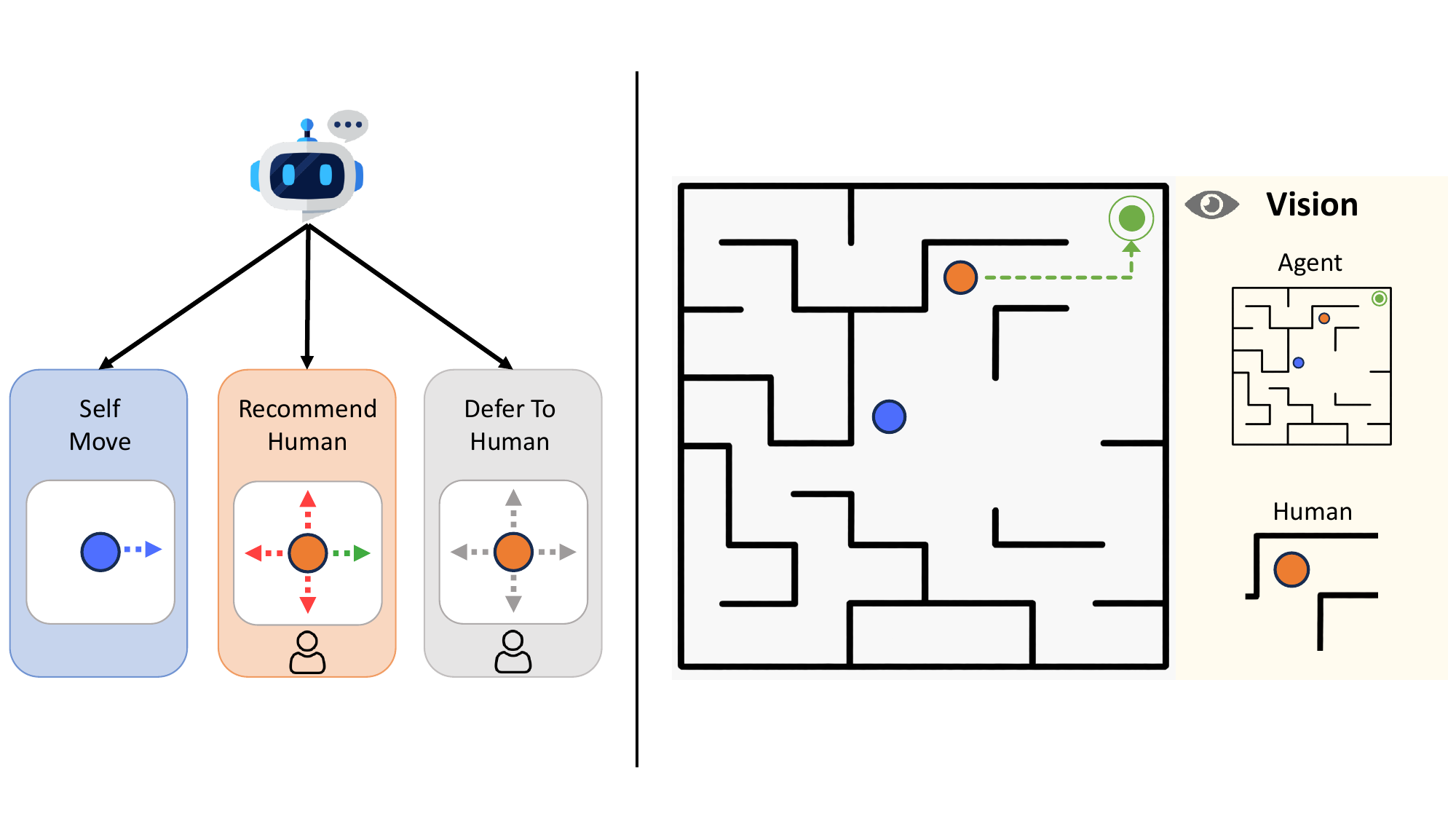}
    \vspace{-10mm}
    \caption{\textbf{Overview of the shared-control maze environment.}
    The agent-controlled avatar (blue) and human-controlled avatar (orange) share the objective of reaching the target (green). At each team turn, the model selects one of three action types: (1)~move its own avatar, (2)~allocate the move to the human with a directional recommendation, or (3)~hand the move over without a recommendation. The model receives the global maze topology and both avatar positions, depicted schematically here, whereas the human acts from a local view. Initial avatar positions vary the relative shortest-path advantage of the agent and human.}
    \vspace{-4mm}
    \label{fig:overview}
\end{figure}

Language-model agents increasingly act alongside people~\citep{lin2024decision,feng2024rehac,shao2026collaborativegym}, and as they grow more capable they can carry more of a task without human assistance.
Falling reliance on a human partner is easy to read as a failure of collaboration, as though the agent were disregarding the person beside it, when it may instead be a correct division of labor.
Collaboration in such settings is a problem of \emph{control allocation}: at each step the agent may act on its own, direct the person to act, or hand the decision over entirely~\citep{scerri2002adjustable,javdani2015shared}.
What makes an allocation good is not how often the person is involved, but whether that involvement responds to the partners' relative advantages on the task at hand~\citep{wilder2020complement,mozannar2020consistent}.
An agent that consults a person who has nothing to contribute wastes a turn, and an agent that acts alone while the person holds the better option wastes an opportunity.
This motivates our central research question:
\begin{quote}
\emph{How does an agent's standalone task capability relate to when and how it involves a human collaborator?}
\end{quote}
%
% TODO(cite): human--agent teaming, mixed-initiative control allocation.

%
Prior work shows that aggregate reliance or human-involvement rates do not by themselves characterize effective human--agent collaboration. We extend this concern to agent-initiated control allocation, where the same involvement rate may reflect either selective use of a better-positioned collaborator or limited independent task capability~\citep{schemmer2023appropriate,bansal2021whole,bucinca2020proxy,shao2026collaborativegym}.
Frequent human involvement may reflect effective allocation to a better-positioned collaborator, but it may equally indicate the agent’s weak independent capability, with the agent outsourcing work it cannot perform itself.
Conversely, low human involvement may reflect competent self-reliance, or a failure to recognize and use a comparatively advantaged partner.
Two agents can exhibit identical aggregate involvement rates while following fundamentally different strategies.
Telling them apart requires two things that aggregate rates lack: a measurement of what the agent accomplishes alone, and a task in which the human's marginal value is known and varied by construction.
We therefore treat standalone capability as a baseline to be measured first, and interpret co-play behavior relative to it.
%
% TODO(cite): human--AI complementarity, help-seeking and ask-human benchmarks.

%
To make the human's marginal value explicit and controllable, we introduce a shared-control maze environment.
An agent-controlled avatar and a human-controlled avatar share the objective of reaching a common target in as few team turns as possible, and either avatar reaching it ends the episode.
Only one avatar moves per team turn, and the model chooses among three actions: (1)~moving its own avatar, (2)~allocating the move to the human together with a directional recommendation, or (3)~handing the move over without one.
This separates using the human as an actor from ceding directional authority to them.
The model observes the full maze topology and both avatar positions, whereas the human acts from a local view.
Initial avatar placement sets the human's positional comparative advantage $\Delta$, the difference between the two shortest-path distances to the target, so that the agent route is shorter, the routes are tied, or the human route is shorter.
Because $\Delta$ is manipulated through geometry alone, the maze, the team objective, and the interaction protocol are held fixed across conditions, and the oracle distances that define $\Delta$ are never stated in the prompt.
The complete system prompts, per-turn user message, and interaction-history extraction procedure are given in Appendix~\ref{app:prompts-history}.
Figure~\ref{fig:overview} summarizes the shared-control environment and the model's available control-allocation actions.

We evaluate GPT~5.6~Sol~\citep{openai2026gpt56}, Command~A+~\citep{cohere2026commandaplus}, Qwen3.8 27B~\citep{qwen2026qwen38}, and Command~R~\citep{coherelabs2024commandr}.
Solo episodes with a single avatar and full-map observability establish that the first three reach near-ceiling performance in this environment, completing essentially every maze with directional optimality at or above $0.97$, whereas Command~R completes $22\%$ of solo episodes with directional optimality of $0.53$.
In co-play, stronger solo performance is associated with selective, route-conditioned human involvement.
The three near-ceiling models select self-action on nearly every initial decision when the routes are tied or favor the agent, and involve the human on $77.5\%$ to $82.5\%$ of initial decisions when the human route is shorter.
Command~R shows no comparable selectivity.
Its whole-game action composition varies comparatively little with which avatar is closer, holding near $35$--$40\%$ self-action and $60$--$65\%$ guidance across relative-advantage states, which is consistent with involving the human relatively indiscriminately rather than in response to comparative advantage.
As a secondary result, human involvement is realized almost entirely through guidance: no model handed over a move without a recommendation on an initial decision in any condition, so models that involved the human retained a role in choosing the direction rather than fully ceding it.
\section{Related Work}
\label{sec:related_work}
Work relevant to our setting converges on a common issue that effective human--AI collaboration depends not only on whether a human participates, but on whether that participation is matched to the task and to the division of control between partners. 
Human--AI complementarity~\citep{bansal2019beyond,wilder2020complement,bansal2021teammate,hemmer2025complementarity} studies how differences in human and machine capabilities can be combined to improve joint performance.
Learning to defer~\citep{madras2018predict,mozannar2020consistent,mozannar2023who,wei2024dependence} makes the allocation problem explicit by learning when a model should produce a prediction itself and when it should pass the decision to a human expert. 
These formulations establish selective allocation as an important part of human--AI performance, but they typically study prediction-time routing under a learned deferral policy. 
In sequential collaboration, control allocation can recur throughout an episode, and the value of involving a partner can change with the current task state.

Appropriate reliance~\citep{schemmer2023appropriate} provides a particularly useful perspective for evaluating such behavior. In AI-assisted decision making, a high reliance rate can reflect appropriate use of accurate advice or over-reliance on incorrect recommendations, so the frequency of reliance is not sufficient to characterize collaboration quality. 
Studies of complementary team performance~\citep{bansal2021whole} show that increasing acceptance of AI recommendations does not necessarily improve team outcomes. Evaluation work on human--AI systems~\citep{bucinca2020proxy} further shows that commonly used aggregate or subjective measures can diverge from objective task performance.
Recent cooperative studies~\citep{gor2026wheel} separate delegation from later adoption of an AI recommendation, highlighting that different forms of participation can have different behavioral meanings. 
Process-oriented human--agent evaluation in Collaborative Gym~\citep{shao2026collaborativegym} similarly examines initiative and control alongside task outcomes. Most of this literature studies how humans rely on AI systems. The corresponding agent-side problem has a related ambiguity: frequent human involvement may indicate effective use of a useful teammate, but it may also reflect limited independent capability; low involvement can reflect competent self-reliance or a failure to exploit a better-positioned partner. This makes involvement rates difficult to interpret without information about both task conditions and what the agent can accomplish on its own.

The distribution of decision authority has also been studied in mixed-initiative interaction~\citep{horvitz1999principles}, adjustable autonomy~\citep{scerri2002adjustable}, and shared autonomy~\citep{javdani2015shared,reddy2018shared}, where control may shift between humans and autonomous systems during a shared task. Human--agent coordination in cooperative environments~\citep{carroll2019utility,strouse2021collaborating} further shows that strong individual or self-play performance does not guarantee effective coordination with human partners. Recent language-model research extends these questions to shared-goal dialogue~\citep{lin2024decision}, long-horizon collaboration~\citep{wu2025collabllm}, and process-oriented evaluation of collaborative agents~\citep{sun2025collabovercooked}. Closely related work also studies selective human intervention. ReHAC~\citep{feng2024rehac} learns when human intervention should occur during complex LLM-agent tasks, while assistance-requesting frameworks~\citep{nguyen2022request,zhang2023goodtime,elfeki2026hilbench} study when agents should seek information or help from humans. These studies establish several mechanisms through which agents can coordinate with, adapt to, or solicit assistance from people. Our evaluation focuses on a complementary question: whether human involvement tracks an experimentally controlled task advantage once the agent's task capability is measured independently. By combining a solo capability baseline with controlled variation in the human teammate's positional comparative advantage, we can distinguish selective, task-conditioned human involvement from involvement associated with weak standalone performance.
\section{Human--Agent Shared Maze}
\label{sec:environment}

We introduce a shared-control maze environment for studying how language models allocate control between themselves and a human partner.
An agent-controlled avatar and a human-controlled avatar share the objective of reaching a common target in
as few team turns as possible.
Either avatar reaching it ends the episode.
On each turn, the model either moves its own avatar or allocates the move to the human.
Varying their initial positions changes the human's task-specific utility while holding the maze and team objective fixed.
We formalize the environment as a two-player cooperative, partially observable, turn-based decision process with asymmetric control.
The global state at turn $t$ is $s_t=(G,g,p_t^{A},p_t^{H},t)$, where $G=(V,E)$ denotes the maze graph, $g \in V$ is the target cell, and $p_t^{A}, p_t^{H} \in V$ are the current positions of the Agent and human avatars, respectively.
Only one avatar can move on each team turn.

\subsection{Maze and State Space}
\label{sec:maze_state}

Each maze is an unweighted grid graph $G=(V,E)$, where vertices are traversable cells and edges are one-step passages between adjacent cells.
Mazes are procedurally generated from fixed random seeds.
The topology and target remain fixed during an episode; only avatar positions change.
The avatars do not block one another and may occupy the same cell.
%
% Generation, target placement, and start sampling are detailed in Appendix~\ref{app:maze-generation}.

\paragraph{Maze generation.}
Each environment instance is a rectangular grid in which every grid cell is a vertex. 
We generate a \emph{perfect maze} using randomized depth-first search with recursive backtracking. 
Beginning from a seed-determined cell, the procedure repeatedly selects an unvisited orthogonally adjacent cell, removes the wall between the two cells, and backtracks whenever no unvisited neighbor remains. 
The resulting undirected graph is a spanning tree over all grid cells.
Consequently, the maze is connected and contains exactly one simple path between every pair of cells.

\paragraph{Target placement.}
The shared target is placed at an endpoint of a maze diameter. 
We first sample a seed-determined source cell, identify a farthest cell from that source using breadth-first search, and then identify a farthest cell from the first endpoint.
One of the two resulting diameter endpoints is selected deterministically from the episode seed. 
Placing the target at a peripheral cell increases the range of feasible start-to-target route lengths.

\paragraph{Exact-distance start sampling.}
After selecting the target $g$, we compute the shortest-path distance $d_G(v,g)$ from every cell $v$ to the target using breadth-first search. 
Given a near-distance parameter $r$ and a nonnegative distance gap $\delta$, candidate start cells are sampled uniformly from the sets
\begin{align}
    V_r &= \{v \in V : d_G(v,g)=r\},\\
    V_{r+\delta} &= \{v \in V : d_G(v,g)=r+\delta\}.
\end{align}
Thus, the realized initial route lengths match the requested distances exactly, rather than approximately. 
When $\delta>0$, the two starts are distinct. 
When
$\delta=0$, both avatars are assigned the same start cell, ensuring identical initial routes as well as identical route lengths. 
Assignment of the near and far cells to the Agent and human avatars is counterbalanced separately in the
experimental design.

Some maze layouts cannot realize a requested pair $(r,r+\delta)$. 
The implementation attempts up to 200 deterministic layouts, beginning with the requested seed and incrementing the seed for each retry. 

% \paragraph{Feasibility and reproducibility.}
%
% An episode is created from the first layout containing cells at both exact distances; if none is found, construction terminates with an error. 
%
% Recording the episode seed and realized configuration makes each accepted instance reproducible.

\subsection{Observation Space and Interaction History}
\label{sec:observation}

The agent receives the complete maze topology, target, and both avatar positions, represented as a textual \emph{adjacency list} that enumerates traversable
connections.
Prior work on LLM maze navigation has found adjacency-list representations to substantially outperform visual grid representations~\citep{li2026spatial}, suggesting that explicit graph structure provides a more reliable interface for LLM spatial planning.
At turn $t$, the agent observation is $o_t^{A}=(\operatorname{Adj}(G),g,p_t^{A},p_t^{H},H_t)$, where $\operatorname{Adj}(G)$ is the adjacency-list representation and $H_t$ is the history through turn $t-1$.
The history is extracted after each resolved interaction as a structured record of the actor, action type, any recommended direction, the executed displacement, and whether a human move followed or overrode a prior recommendation.
This log is rendered as text and inserted into the next model prompt.
The full prompts and extraction procedure are given in Appendix~\ref{app:prompts-history}.
The human instead acts from a local view centered on their avatar and may follow or override the model's recommendation.

\subsection{Action Space}
\label{sec:action_space}

At every team turn, the language model makes the control-allocation decision.
Let $\mathcal{D}=\{\texttt{UP},\texttt{DOWN},\texttt{LEFT},\texttt{RIGHT}\}$ denote the set of movement directions.
There are three types of action for the model to choose:
(1)~Self-action: With $\operatorname{MOVE\_SELF}(d)$, the model moves its avatar one cell in direction $d$.
(2)~Guided human action: With $\operatorname{GUIDE\_HUMAN}(d)$, the model allocates the move to the human while recommending $d$, which the human may follow or override.
(3)~Human deference: With $\operatorname{DEFER\_TO\_HUMAN}$, the model allocates the move without a recommendation, leaving the direction entirely to the human.
The model selects from $\mathcal{A}^{A}=\{\operatorname{MOVE\_SELF}(d),
\operatorname{GUIDE\_HUMAN}(d),\operatorname{DEFER\_TO\_HUMAN}\}$, where
$d\in\mathcal{D}$.
This distinguishes using the human avatar from fully ceding directional
authority.

\subsection{Task-Conditioned Comparative Advantage}
\label{sec:comparative_advantage}

The environment controls the relative utility of the human through the avatars' positions in the maze.
Let $d_t^{A}=d_G(p_t^{A},g)$ and $d_t^{H}=d_G(p_t^{H},g)$ denote the avatars' shortest-path distances to the target.
At the beginning of an episode, we define the human's positional comparative advantage as $\Delta=d_0^{A}-d_0^{H}$.
Thus we have three different situations: (1)~$\Delta > 0$ indicates that the human-controlled avatar has the shorter route to the target, (2)~$\Delta < 0$ indicates that the agent-controlled avatar has the shorter route, and (3)~$\Delta = 0$ indicates equal shortest-path distances.
$\Delta$ changes the utility of involving the human without changing the team objective or interaction protocol.
These oracle distances and $\Delta$ are used for construction and evaluation but are not stated in the model prompt.
The model may only infer them from the maze and avatar positions.
The specific values of $\Delta$ used in our study are described in Section~\ref{sec:experiments} later.

\subsection{Transition Dynamics and Termination}
\label{sec:transition}

Each team turn produces at most one avatar movement.
A $\operatorname{MOVE\_SELF}(d)$ action immediately executes the agent move.
Under $\operatorname{GUIDE\_HUMAN}(d)$ or $\operatorname{DEFER\_TO\_HUMAN}$, the human's selected movement completes the turn.
The resolved state is then added to the history before the next model decision.
The episode ends successfully when either avatar reaches the target, or unsuccessfully when the maximum horizon $T_{\max}$ is reached.
\section{Experiments}
\label{sec:experiments}

\subsection{Model Capability Baseline}
\label{sec:capability-baseline}

We evaluate Qwen~3.8~27B, Command~R, Command~A+, and GPT~5.6~Sol.
Additional model-access, licensing, inference-setting, and compute-resource
details are provided in Appendix~\ref{app:model_details}.
Because model identity is not itself a capability measure, we separately evaluate each model in solo episodes with one avatar and full-map observability.
This baseline provides the reference point needed to interpret whether subsequent co-play behavior reflects selective human involvement or limited independent task capability.
For comparability with co-play, Table~\ref{tab:solo-capability} uses the same adjacency-list representation, interaction-history condition, and prompt family (Appendix~\ref{app:prompts-history}) for every model.
Each model completed 50 matched-seed $5\times5$ mazes with shortest path length~$5$ and a ten-turn horizon. 
Let $C$ denote completion turn and $d_{ai}$ the model's initial BFS distance to the target. S@$k$ is the fraction of episodes with $C\leq k$, evaluated at the checkpoint ($k=7$) and horizon ($k=10$). 
Normalized efficiency is $E=d_{ai}/C$ for success by turn 10 and zero for failure. 
Model directional optimality (MDO) is the fraction of proposed directions that reduce BFS distance by one. 
Paired seed-bootstrap confidence intervals and parse/retry diagnostics appear in Appendix~\ref{app:metrics-results}, Tables~\ref{tab:solo-capability-full} and~\ref{tab:solo-diagnostics}.

\begin{table}[t]
  \centering
  \caption{Task-specific capability on solo $5\times5$ mazes, with $N=50$
  matched seeds per model.}
  \label{tab:solo-capability}
  \small
  \setlength{\tabcolsep}{15pt}
  \begin{tabular}{@{}lccccc@{}}
    \toprule
    Model & $N$ & S@7 & S@10 & $E$ & MDO \\
    \midrule
    Qwen~3.8~27B & 50 & 0.98 & 0.98 & 0.97 & 0.98 \\
    Command~A+   & 50 & 0.96 & 1.00 & 0.96 & 0.97 \\
    Command~R    & 50 & 0.18 & 0.22 & 0.20 & 0.53 \\
    GPT~5.6~Sol  & 50 & 1.00 & 1.00 & 1.00 & 1.00 \\
    \bottomrule
  \end{tabular}
\end{table}

GPT~5.6~Sol finished every maze on the oracle path. 
Command~A+ also completed every episode, with a small efficiency cost from occasional detours, and Qwen-3.8 27B timed out in only one episode.
These three models therefore form a near-ceiling-capability group in this environment. 
By contrast, Command~R completed 22\% of episodes, with MDO $0.53$, providing a substantially lower-capability comparison. 
This distinction motivates examining whether human involvement is allocated differently depending on the agent's own ability to solve the task. 
Co-play results involving Command~R should therefore be interpreted jointly with its limited navigation competence rather than as allocation behavior alone. 
With only four evaluated models, these groups support a descriptive comparison, not a continuous scaling relationship between capability and collaboration.

\subsection{Co-play Design and Measures}
\label{sec:measures}
Four of the paper's authors served as human operators in the co-play evaluation.
Each completed 120 episodes, yielding 480 episodes in total. 
Participants were instructed to reach the shared target in as few team turns as possible. 
The instruction did not reward involving either avatar, following guidance, or ceding control. 
We cross model, participant, ten fixed maze seeds, and initial route advantage $\Delta_0\in\{-2,0,+2\}$, yielding 40 episodes per model--advantage cell.
The $5\times5$ mazes have a shortest path of five moves, so the initial route pairs are $(5,7)$, $(5,5)$, and $(7,5)$ for Agent-favored, tied, and human-favored episodes.
Each episode has a ten-turn horizon.

Our primary outcome is the first control-allocation decision. 
For $a\in\{S,G,D\}$, denoting self-action, guidance, and deference, respectively, we write the conditional first-action probability as \(P_m^a(\delta)=\Pr(a_{i1}=a\mid m_i=m,\Delta_{i0}=\delta)\).
The initial self-action, guidance, and deference rates are \(\mathrm{ISR}_m(\delta)=P_m^S(\delta)\), \(\mathrm{IGR}_m(\delta)=P_m^G(\delta)\), and \(\mathrm{IDR}_m(\delta)=P_m^D(\delta)\).
Our claim-focused summaries are equal-route self-action \(\mathrm{ZSR}_m=\mathrm{ISR}_m(0)\), equal-route self-action bias \(\mathrm{ZSB}_m=\mathrm{ZSR}_m-0.5\), non-human-favored self-action \(\mathrm{NFSR}_m=[\mathrm{ISR}_m(-2)+\mathrm{ISR}_m(0)]/2\), human-favored uptake \(\mathrm{HFU}_m=1-\mathrm{ISR}_m(+2)\), and self-action selectivity \(\mathrm{SSC}_m=\mathrm{NFSR}_m-\mathrm{ISR}_m(+2)\).
Secondary measures capture actor-aligned allocation, directional optimality, and team success and efficiency. 
Appendix~\ref{app:metrics-results}, Table~\ref{tab:metric-reference}, provides the complete definitions.
Detailed results and confidence intervals appear in Appendix~\ref{app:initial-control}--\ref{app:whole-game}, Tables~\ref{tab:initial-actions-full}--\ref{tab:team-outcomes-full}. 
We compute 95\% percentile intervals using 1,000 two-way cluster-bootstrap samples over participants and maze seeds. 
Given four participants, all comparisons are descriptive rather than population-level estimates.

\begin{table}[t]
  \centering
  \caption{Initial control allocation. Both panels report point estimates. Positive $\Delta_0$ means that the human has the shorter route.}
  \label{tab:main-allocation}
  \small
  \begin{minipage}{\linewidth}
  \centering
  \textbf{Panel A: observed first actions}\par
  \vspace{3pt}
  \setlength{\tabcolsep}{15pt}
  \begin{tabular}{@{}lrcccc@{}}
    \toprule
    Model & $\Delta_0$ & $N$ & Self (ISR) & Guide (IGR) & Defer (IDR) \\
    \midrule
    Qwen~3.8~27B & $-2$ & 40 & 1.000 & 0.000 & 0.000 \\
                  & $0$  & 40 & 1.000 & 0.000 & 0.000 \\
                  & $+2$ & 40 & 0.175 & 0.825 & 0.000 \\
    \addlinespace
    Command~A+   & $-2$ & 40 & 0.950 & 0.050 & 0.000 \\
                  & $0$  & 40 & 0.975 & 0.025 & 0.000 \\
                  & $+2$ & 40 & 0.225 & 0.775 & 0.000 \\
    \addlinespace
    Command~R    & $-2$ & 40 & 0.700 & 0.300 & 0.000 \\
                  & $0$  & 40 & 0.450 & 0.550 & 0.000 \\
                  & $+2$ & 40 & 0.575 & 0.425 & 0.000 \\
    \addlinespace
    GPT~5.6~Sol  & $-2$ & 40 & 1.000 & 0.000 & 0.000 \\
                  & $0$  & 40 & 1.000 & 0.000 & 0.000 \\
                  & $+2$ & 40 & 0.175 & 0.825 & 0.000 \\
    \bottomrule
  \end{tabular}

  \par\vspace{8pt}
  \textbf{Panel B: claim-focused summaries}\par
  \vspace{3pt}
  \setlength{\tabcolsep}{4pt}
  \tablefit{%
  \begin{tabular}{@{}lccccc@{}}
    \toprule
    Model &
    \shortstack{Equal-route self\\ZSR} &
    \shortstack{Bias from 0.5\\ZSB} &
    \shortstack{Self when human\\not favored (NFSR)} &
    \shortstack{Human-favored uptake\\HFU} &
    \shortstack{Self-action selectivity\\SSC} \\
    \midrule
    Qwen~3.8~27B & 1.000 & 0.500 & 1.000 & 0.825 & 0.825 \\
    Command~A+   & 0.975 & 0.475 & 0.963 & 0.775 & 0.738 \\
    Command~R    & 0.450 & $-$0.050 & 0.575 & 0.425 & 0.000 \\
    GPT~5.6~Sol  & 1.000 & 0.500 & 1.000 & 0.825 & 0.825 \\
    \bottomrule
  \end{tabular}%
  }
  \end{minipage}
\end{table}

\subsection{Results}
\label{sec:results}

\begin{figure}[!htbp]
  \centering
  \includegraphics[width=\linewidth]{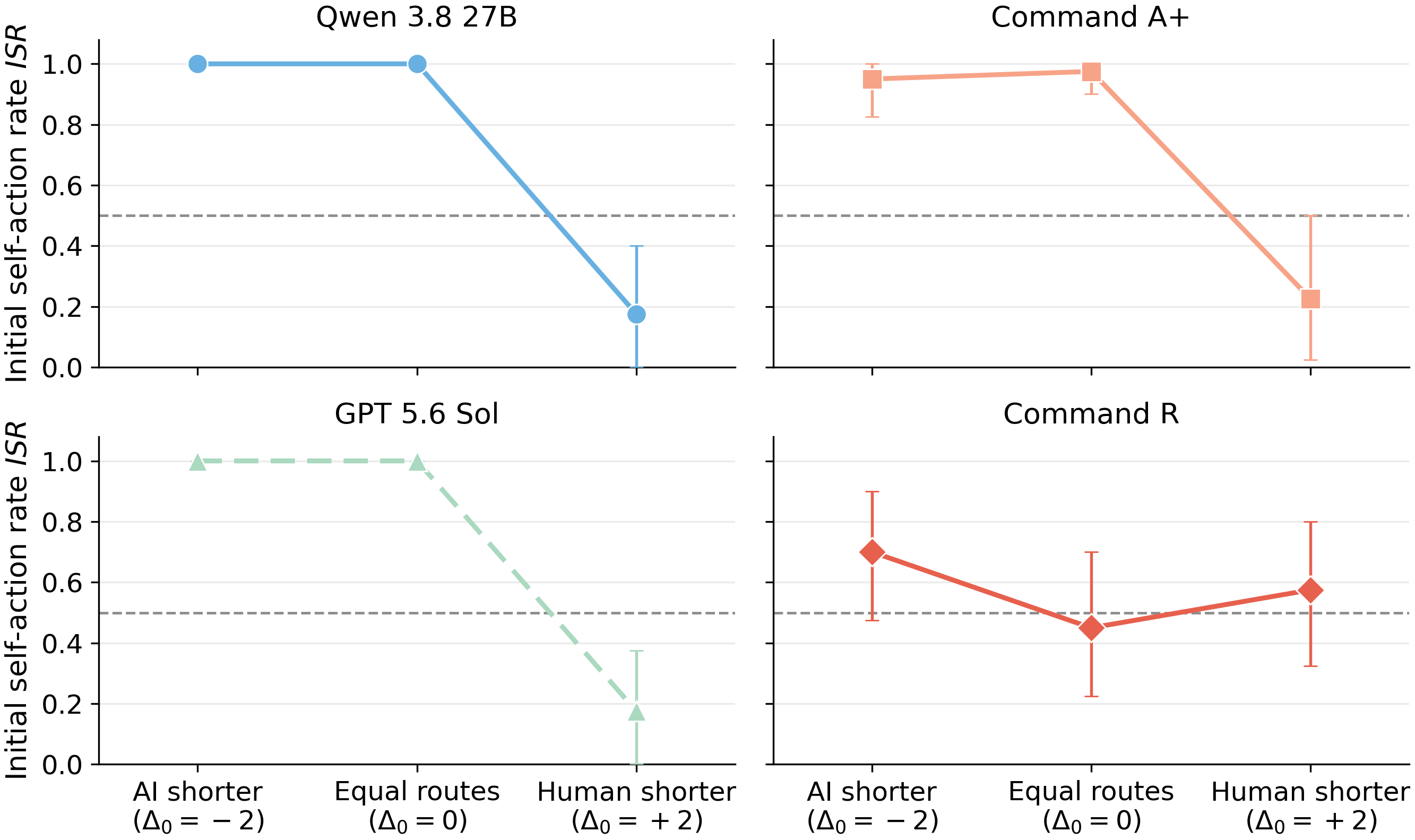}
  \caption{\textbf{Initial self-action by assigned route advantage.} Points show the initial self-action rate (ISR) under AI-shorter ($\Delta_0=-2$), equal-route ($\Delta_0=0$), and human-shorter ($\Delta_0=+2$) starting conditions. Error bars show 95\% two-way cluster-bootstrap confidence intervals over participants and maze seeds. Each model--condition cell contains 40 episodes. The dashed horizontal line marks a descriptive reference value of 0.5.}
  \label{fig:initial-self-action}
\end{figure}

\paragraph{Control allocation tracks comparative advantage for three models.}
Table~\ref{tab:main-allocation} reports the observed first actions and the claim-focused summaries, and Figure~\ref{fig:initial-self-action} visualizes the initial self-action rates across assigned advantage conditions. 
Both table
panels present point estimates; their full confidence intervals are reported in
Appendix~\ref{app:initial-control},
Tables~\ref{tab:initial-actions-full} and \ref{tab:initial-claims-full}.

Qwen~3.8~27B and GPT~5.6~Sol always selected self-action when the AI route was shorter or tied, but involved the human on 82.5\% of initial decisions when the human route was shorter. 
Command~A+ exhibited a similar but weaker shift, selecting self-action on 97.5\% of tied-route decisions and involving the human on 77.5\% of human-favored decisions. 
Command~R did not exhibit positive self-action selectivity: its initial self-action rate was lower when the routes were tied than when the human was favored. 
%
%Human involvement occurred overwhelmingly through guidance rather than full deference: no model selected deference on an initial decision in any condition.

%
Taken together, the observed results show that the three models with near-ceiling solo capability almost always selected self-action when the initial routes were tied, but substantially increased human involvement when the human had the shorter route. 
This condition-dependent contrast yields positive SSC values for Qwen~3.8~27B, Command~A+, and GPT~5.6~Sol, whereas Command~R has an SSC of approximately zero. 
We interpret this pattern as evidence that, within the four evaluated models, stronger standalone task capability is associated with more selective, task-conditioned human involvement rather than uniformly more or uniformly less human involvement. 
The comparison is associational and does not establish that increasing capability causes more selective allocation.

\paragraph{Allocation quality and task performance diverge by model.}
Qwen~3.8~27B and GPT~5.6~Sol completed every episode by turn ten across all three advantage conditions.
Command~A+ completed 97.5\%, 100\%, and 87.5\% of episodes in the AI-favored, tied, and human-favored conditions, respectively.
Command~R completed only 17.5\%, 22.5\%, and 45.0\%, although its performance improved when the human held the shorter route.
Thus, Command~R achieved higher team completion when the human-controlled avatar began closer. 
Notably, this performance improvement occurred without a corresponding condition-specific increase in Command R’s whole-game human-involvement rate.
However, the experiment assigns initial route advantage rather than human involvement itself, so this comparison does not identify a causal effect of increased human involvement on completion.
The directional-optimality results in Appendix~\ref{app:whole-game}, Table~\ref{tab:whole-game-full}, show that these allocation patterns must be interpreted jointly with each model's navigation competence rather than as collaboration behavior alone.

\paragraph{Human involvement occurs primarily through guidance.}
Human involvement occurred overwhelmingly through guidance rather than full deference: no model selected deference on an initial decision in any condition.
The whole-game results show the same pattern, with deference shares of zero or near zero across models and conditions (Appendix~\ref{app:whole-game}, Table~\ref{tab:whole-game-full}).
Thus, when models involved the human, they generally allocated the move while retaining a role in direction selection through a recommendation, rather than fully ceding directional choice. 
This secondary result complements the primary association between solo capability and allocation selectivity.

\begin{figure}[!htbp]
  \centering
  \includegraphics[width=\linewidth]{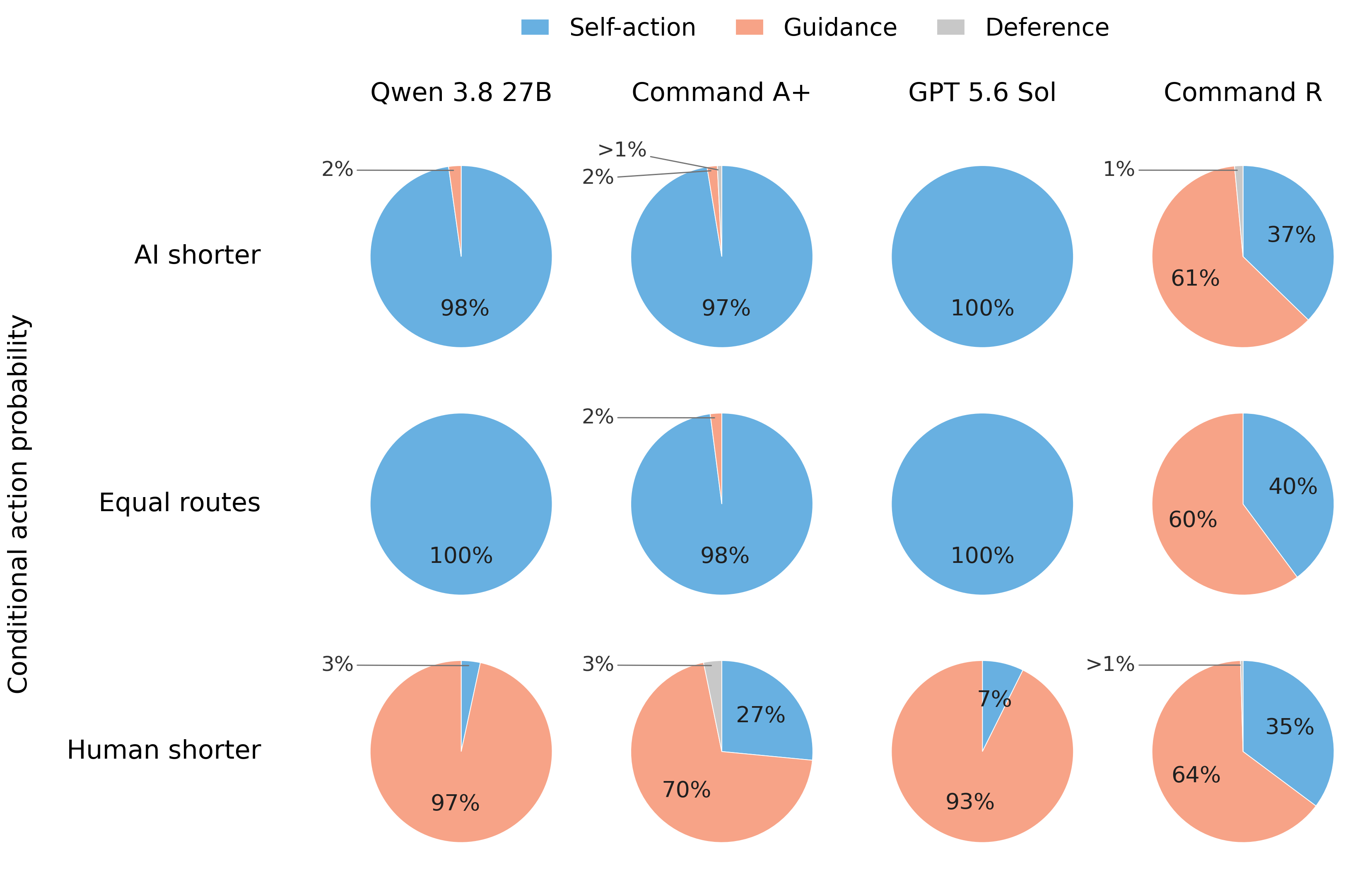}
  \caption{\textbf{Action allocation by current relative route advantage.} Columns correspond to models, and rows group observed turns according to whether the Agent-controlled avatar has the shorter current route, the two routes are equal, or the human-controlled avatar has the shorter current route. Pie slices show the Current-state action rate (CAR) of self-action, guidance, and deference within each model--advantage cell. Current route advantage is evaluated immediately before each model decision.}
  \label{fig:current-action-allocation}
\end{figure}

\paragraph{Whole-game action composition varies by model.}
Figure~\ref{fig:current-action-allocation} reports action composition conditional on current relative route advantage across all observed turns.
Qwen~3.8~27B, Command~A+, and GPT~5.6~Sol primarily selected self-action when the Agent-controlled avatar was closer or the routes were tied, but shifted substantially toward guidance when the human-controlled avatar was closer.
By contrast, Command~R's action composition varied comparatively little across current-advantage states, remaining near 35--40\% self-action and 60--65\% guidance regardless of which avatar was closer. 
This descriptive pattern is consistent with less task-conditioned allocation by Command~R. 
Because current route advantage may be affected by preceding actions, these whole-game comparisons are descriptive rather than causal.

% Let concluding text share space with the results figures.

\section{Conclusion and Discussion}

We studied how an AI agent's standalone task capability relates to when and how it involves a human collaborator.
The solo baseline revealed a substantial capability difference within the four evaluated models: GPT~5.6~Sol, Command~A+, and Qwen~3.8~27B achieved near-ceiling performance in the maze environment, whereas Command~R completed substantially fewer solo episodes.
This baseline is essential for interpreting the co-play results because the same observed rate of human involvement may reflect either selective allocation or limited independent task capability.

The models with near-ceiling independent capability exhibited selective human involvement conditioned on relative route advantage. 
They predominantly selected self-action when the AI route was shorter or tied, but substantially increased human involvement when the human had the shorter route. 
These condition-dependent shifts produced positive self-action selectivity scores.
Command~R, which had substantially weaker solo performance, did not exhibit the same positive allocation pattern. 
Across the evaluated models, greater indepedent task capability was therefore associated with more selective, task-conditioned human involvement, rather than uniformly more or uniformly less collaboration. 
When models involved the human, they did so primarily through guidance rather than full deference, allocating the move while usually providing a directional recommendation.

Consistent with prior work showing that collaboration depends on agent capability and on the form and timing of human participation, our findings demonstrate how aggregate involvement rates can obscure allocation selectivity in sequential shared control. 
Capability baselines and relative-advantage-conditioned allocation measures provide complementary information about how control is distributed.
Aggregate human-involvement rates alone can obscure whether an agent is responding to the human's comparative advantage or involving the human despite being unable to perform the task reliably itself. 
Capability baselines and condition-specific allocation measures therefore provide complementary information about how control is distributed in collaborative tasks.

\section{Limitations and Future Work}

Our conclusions are limited by the evaluation of only four models, the small number of human participants, and the use of a single maze-based environment and interaction protocol. 
The resulting comparisons are descriptive of the evaluated models and setting. 
In particular, they establish an association between stronger solo performance and selective human involvement.
Establishing the generality and direction of this relationship will require broader model samples, participant populations, task domains, and collaboration protocols.

Future work will extend the same capability-conditioned analysis beyond navigation to settings where the human's comparative advantage arises from other sources, including cooperative game environments, everyday work tasks, and collaborative coding.
Because the measures used here depend only on a solo capability baseline and a controlled variation in the partner's task-specific advantage, they transfer to any shared task in which that advantage can be manipulated by construction.
\section{Ethical Considerations}

The purpose of this study is to better understand and evaluate the collaborative behavior of language-model agents, with the broader goal of informing the development of AI systems that collaborate with humans more effectively.
Our results should not be interpreted as advocating for the exclusion of humans from tasks when an AI system is capable of acting independently, nor as suggesting that greater model capability necessarily reduces the value of human participation.
Rather, we study whether agents adapt their involvement of a human collaborator according to the collaborator's task-specific advantage.
Understanding such behavior may help identify when capable AI systems appropriately make use of human knowledge, judgment, or abilities, and when their allocation of control could be made more responsive to their human collaborators.

The shared-control maze is a simplified experimental environment designed to study these behaviors under controlled conditions.
Our findings therefore should not be used to justify reducing human oversight or decision-making authority in consequential real-world settings.

\section*{Acknowledgments}

This research was supported in part by Lambda, Inc., which provided cloud GPU computing credits that enabled self-hosted model inference, and by Cohere Inc., which provided research grant support for API access to its language models. These resources were instrumental in conducting our experiments and systematically evaluating how model capability and task conditions influence human--AI collaboration.

%%%%%%%%%%%%%%%%%%%%%%%%%%%%%%%%%%%%%%%%%%%%%%%%%%%%%%%%%%%%
\newpage
\bibliographystyle{unsrt}
\bibliography{reference_refined}

%%%%%%%%%%%%%%%%%%%%%%%%%%%%%%%%%%%%%%%%%%%%%%%%%%%%%%%%%%%%
\newpage
\appendix
\raggedbottom
\section{Prompts and History Extraction}
\label{app:prompts-history}

This appendix records the prompts used in the reported experiments and the
procedure that extracts interaction history for the next model call.
Solo capability evaluation uses the solo prompt family
(Listings~\ref{lst:solo-system}--\ref{lst:solo-output}); co-play uses the
shared-control prompt family
(Listings~\ref{lst:coplay-system}--\ref{lst:coplay-output}).
Both conditions use the adjacency-list state rendering of
Li et al.~\citep{li2026spatial} and include interaction history at every turn.
The partner is always referred to as a teammate; the prompt never identifies
the partner as a human, never states oracle distances or $\Delta$, and never
discloses the remaining turn budget.

\subsection{Observation Extraction}
\label{app:observation-extraction}

Before each model call, a \emph{neutral observation} is extracted from the live
environment. It contains only (i)~the full wall grid, (ii)~the target cell,
(iii)~both avatar positions, and (iv)~the legal-move sets of the acting avatar
and, in co-play, of the teammate. Oracle shortest-path distances, $\Delta$,
and any statement of who is closer are excluded from this object, so they
cannot enter the prompt.

The observation is rendered as tagged adjacency-list text. Coordinates are
explained once, then the maze is given as an undirected edge list of open
passages, followed by the model's cell (\texttt{ORIGIN}), the target
(\texttt{TARGET}), the teammate cell when present, and both legal-move sets:
\begin{lstlisting}[caption={Adjacency-list state block (schematic). Legal-move lists are instance-specific.},label=lst:adj-state]
Coordinates are (x, y): x is the column (0 = left), y is the row
(0 = top, increasing downward). Moves are U/D/L/R, not compass letters:
U = up (y-1), D = down (y+1), L = left (x-1), R = right (x+1).
<ADJLIST_START> (0,0) <--> (1,0) ; (1,0) <--> (1,1) <ADJLIST_END>
<ORIGIN_START> (0,0) <ORIGIN_END>
<TARGET_START> (4,1) <TARGET_END>
<TEAMMATE_START> (2,3) <TEAMMATE_END>
Your legal moves: [R, D]
Teammate legal moves: [U, L]
\end{lstlisting}

\subsection{History Extraction}
\label{app:history-extraction}

Let $H_t$ denote the interaction history available at the start of team
turn~$t$, with $H_1=\emptyset$. After each resolved interaction, a structured
record is appended. Model decisions contribute one of the following:
\begin{itemize}
    \item Self-action: actor, start cell, end cell, direction, and whether the
    move passed through an open passage.
    \item Guidance: actor, recommended direction, and whether that direction
    was legal for the teammate at the time of recommendation.
    \item Deference: actor and the deferral tag, with no direction.
\end{itemize}
A subsequent human move contributes actor, start cell, end cell, direction,
validity, the recommendation that was pending when the move was issued (if
any), and a follow/override bit. The pending recommendation is captured
\emph{before} the move is applied, so the next prompt can show whether the
human followed or overrode the advice.

At the next model call, $H_t$ is rendered as a numbered text log and inserted
under the heading ``Interaction history:''. Empty history is shown as
\texttt{(no moves yet)}. Actor labels are rewritten from the model's
perspective (\texttt{you} vs.\ \texttt{teammate}). Guidance and deference
appear as allocation lines; executed movements appear as coordinate
transitions. When a teammate move resolves a prior recommendation, the move
line is annotated with follow or override. Listing~\ref{lst:history-example}
shows a compact example covering guidance followed, guidance overridden, and
deference.

\begin{lstlisting}[caption={Rendered interaction history covering guidance, follow/override, and deference.},label=lst:history-example]
Interaction history:
Step 1 [you]: GUIDE_TEAMMATE R (a suggestion; teammate may follow or override)
Step 2 [teammate]: (0,0) --R--> (1,0) [valid] (followed your R suggestion)
Step 3 [you]: GUIDE_TEAMMATE U (a suggestion; teammate may follow or override)
Step 4 [teammate]: (1,0) --D--> (1,1) [valid] (you suggested U; overrode with D)
Step 5 [you]: DEFER (handed the turn over, no suggestion)
Step 6 [teammate]: (1,1) --R--> (2,1) [valid]
\end{lstlisting}

The per-turn user message concatenates the rendered state, the history block,
and the output-contract instructions below. If the model reply is unparseable
or proposes an illegal self-move or illegal recommendation, a clarification
turn restates the output contract and the relevant legal-move set; the
original system prompt is unchanged.

\subsection{Solo Prompts}
\label{app:solo-prompts}

\begin{lstlisting}[caption={Solo system prompt.},label=lst:solo-system]
You are an expert maze solver controlling one avatar on a 2D grid. The objective is to reach the exit cell using as few moves as possible; reaching the exit finishes the task. You can see the entire maze layout. Each turn you choose exactly one move to an orthogonally adjacent cell, and you may only move through an open passage -- never through a wall. Moves are U (up), D (down), L (left), and R (right). Reason step by step about the shortest route to the exit, then commit to a single move. Always end your reply with a line of the form 'MOVE: <DIR>' where <DIR> is U, D, L, or R.
\end{lstlisting}

\begin{lstlisting}[caption={Solo per-turn output instructions. The bracketed list is replaced by the avatar's legal moves.},label=lst:solo-output]
Think step by step about the shortest route from the origin to the target along the passages above. Then, on the final line, output exactly one move for your avatar in the form 'MOVE: <DIR>', where <DIR> is one of [U, R, D].
\end{lstlisting}

\subsection{Co-play Prompts}
\label{app:coplay-prompts}

\begin{lstlisting}[caption={Co-play system prompt.},label=lst:coplay-system]
You and a teammate are navigating one shared maze together toward a single shared exit. You each control one avatar on the grid. As soon as either avatar -- yours or your teammate's -- reaches the exit, the team's task is finished, so your goal is to get an avatar to the exit in as few total team turns as possible. Only one avatar moves on each turn.

You control only your own avatar. You do NOT control your teammate: you cannot move their avatar for them and cannot force what they do. You can see the entire maze layout; your teammate cannot -- they only see a small local area around their own avatar and have no view of the rest of the maze.

The two avatars never block each other. Only walls restrict movement: both avatars may occupy the same cell at the same time, either avatar may move into or through the cell the other is standing on, and neither one ever has to wait for or go around the other. Plan each route as if the other avatar were not on the grid at all.

On each of your turns you choose exactly one of three actions:
- MOVE_SELF <DIR>: move your own avatar one cell through an open passage.
- GUIDE_TEAMMATE <DIR>: recommend a direction to your teammate and hand them the turn. They are free to follow your recommendation or to make a different legal move instead. Use it when you want your teammate to act this turn and you have advice that may help them.
- DEFER: hand the turn to your teammate with no recommendation.

<DIR> is U (up), D (down), L (left), and R (right).

Reason step by step, then commit to exactly one action. Always end your reply with a single final line of one of these forms: 'ACTION: MOVE_SELF <DIR>', 'ACTION: GUIDE_TEAMMATE <DIR>', or 'ACTION: DEFER'.
\end{lstlisting}

\begin{lstlisting}[caption={Co-play per-turn output instructions. Bracketed lists are replaced by the legal-move sets.},label=lst:coplay-output]
Decide your single action for this turn. Think step by step about how the team can reach the exit in as few turns as possible, then end with exactly one final line, one of:
- 'ACTION: MOVE_SELF <DIR>' where <DIR> is one of [U, R, D] (your own legal moves), or
- 'ACTION: GUIDE_TEAMMATE <DIR>' to recommend a direction to your teammate -- a suggestion only, which they may follow or override with another legal move, since you do not control their avatar, or
- 'ACTION: DEFER' to hand your teammate the turn with no recommendation.
\end{lstlisting}
\clearpage
\section{Model Access, Licensing, and Compute}
\label{app:model_details}

\subsection{Model Access and Licensing}

We evaluate GPT~5.6~Sol, Command~A+, Qwen~3.8~27B, and Command~R.
GPT~5.6~Sol is a proprietary model accessed through the OpenAI API and is governed by the applicable OpenAI service terms.
The open-weight releases of Command~A+ and Qwen~3.8~27B are licensed under Apache~2.0.
Command~R 08-2024 is released under CC-BY-NC~4.0 together with the Cohere Labs Acceptable Use Addendum.
All models were used subject to the applicable licenses and provider terms.

\subsection{Model Inference Settings and Compute Resources}

Table~\ref{tab:compute} summarizes the inference configuration and approximate
compute expenditure for all reported solo and co-play experiments. For models
accessed through hosted APIs, provider-side hardware specifications are not
available to us. Reported costs reflect model inference or compute rental costs.

\begin{table}[H]
\centering
\small
\caption{Model inference settings and approximate compute expenditure across all
reported experiments. The Cohere cost is the combined expenditure for
Command~R and Command~A+.}
\label{tab:compute}
\footnotesize
\setlength{\tabcolsep}{4pt}
\tablefit{%
\begin{tabular}{lllll}
\toprule
\textbf{Model} &
\textbf{Access} &
\textbf{Reasoning setting} &
\textbf{Compute / runtime} &
\textbf{Approx. cost} \\
\midrule
Command~R
& Cohere API
& No CoT
& Provider-managed
& $<\$1$ \\

Command~A+
& Cohere API
& No effort setting
& Provider-managed
& $<\$1$ \\

GPT~5.6~Sol
& OpenAI API
& Medium
& Provider-managed
& $<\$10$ \\

Qwen~3.8~27B
& Self-hosted
& Medium
& $1\times$ H100, $\sim$31 GPU-hours
& $\sim\$102$ \\
\bottomrule
\end{tabular}
}
\end{table}
\clearpage
\section{Metric Reference and Detailed Results}
\label{app:metrics-results}

Table~\ref{tab:metric-reference} expands the abbreviations used in the main
text and result tables. Initial rates are calculated across episodes using the
first model decision, whereas whole-game shares are first calculated within
each episode and then averaged across episodes. In the solo capability
evaluation, S@$k$, RCT10, $E$, and MDO are computed on the model's single
avatar; $E$ uses the model's oracle route length $d_{ai}$.

\begin{table}[H]
  \centering
  \caption{Reference for the symbols and abbreviations used in the evaluation.
  ``Human use'' includes both guided and unguided human turns.}
  \label{tab:metric-reference}
  \footnotesize
  \setlength{\tabcolsep}{3pt}
  \begin{tabular}{@{}p{1.05cm}p{2.75cm}p{4.45cm}p{4.55cm}@{}}
    \toprule
    Term & Full name & Definition & Interpretation \\
    \midrule
    $\Delta_0$
      & Initial route advantage
      & $d_0^A-d_0^H$.
      & Positive values mean that the human starts closer; negative values mean
        that the AI starts closer. \\
    ISR
      & Initial self-action rate
      & Fraction of episodes whose first action is
        $\operatorname{MOVE\_SELF}$.
      & How often the model initially chooses its own avatar. \\
    IGR
      & Initial guidance rate
      & Fraction of episodes whose first action is
        $\operatorname{GUIDE\_HUMAN}$.
      & How often the model initially uses the human while retaining
        directional influence. \\
    IDR
      & Initial deference rate
      & Fraction of episodes whose first action is
        $\operatorname{DEFER\_TO\_HUMAN}$.
      & How often the model initially gives the human a turn without advice. \\
    ZSR
      & Equal-route self-action rate
      & $\mathrm{ISR}(0)$.
      & Self-action when neither avatar has a route advantage. \\
    ZSB
      & Equal-route self-action bias
      & $\mathrm{ZSR}-0.5$.
      & Positive values indicate more self-action than both human-involving
        actions combined at a tie. \\
    NFSR
      & Non-human-favored self-action rate
      & $[\mathrm{ISR}(-2)+\mathrm{ISR}(0)]/2$.
      & Self-action when the human has no initial route advantage. \\
    HFU
      & Human-favored uptake
      & $1-\mathrm{ISR}(+2)$.
      & How often the model uses the human when the human starts closer. \\
    SSC
      & Self-action selectivity
      & $\mathrm{NFSR}-\mathrm{ISR}(+2)$.
      & Positive values indicate a calibrated shift from self-action toward the
        human as the human becomes favored. \\
    $q^S,q^G,q^D$
      & Whole-game action shares
      & Within-episode proportions of self, guide, and defer actions.
      & Whether the initial allocation pattern persists across the episode. \\
    AAR
      & Actor-aligned allocation rate
      & Fraction of non-tied turns assigned to the avatar with lower current
        oracle distance.
      & Quality of the model's actor choice, independent of direction. \\
    MDO
      & Model directional optimality
      & Fraction of proposed directions that reduce the selected avatar's
        oracle distance by one.
      & Navigation quality conditional on proposing a direction. \\
    S@$k$
      & Success by turn $k$
      & Fraction of episodes completed by turn $k$.
      & Completion at the fast ($k=7$) or full ($k=10$) cutoff. \\
    RCT10
      & Restricted completion time
      & Completion turn for successes and 10 for episodes unfinished at turn 10.
      & Censoring-aware descriptive completion time; lower is better. \\
    $E$
      & Normalized efficiency
      & $\min(d_0^A,d_0^H)/C$ in co-play, or $d_{ai}/C$ in solo, for success
        by turn 10 and zero otherwise.
      & Success and speed relative to the initially available oracle route. \\
    CAR
      & Current-state action rate
      & Action probability conditional on whether the AI, neither avatar, or
        the human is currently closer.
      & Descriptive adaptation as relative positions change during play. \\
    \bottomrule
  \end{tabular}
\end{table}

\subsection{Solo Capability}
\label{app:solo-capability}

Table~\ref{tab:solo-capability-full} reports seed-bootstrap 95\% percentile
intervals for the point estimates in Table~\ref{tab:solo-capability}. Maze
seed is the resampling unit (1,000 replicates); the four models share the
same 50 seeds, so pairing is preserved. Horizon timeouts
(\texttt{done=true}, \texttt{completed\_by=null}) are valid failures.
Table~\ref{tab:solo-diagnostics} reports invalid-move, unparseable, and retry
rates over the 1,230 recorded LLM steps.

\begin{table}[H]
  \centering
  \caption{Solo capability with seed-bootstrap 95\% confidence intervals.
  Each model has 50 matched-seed episodes. Lower RCT10 and higher $E$ are
  better.}
  \label{tab:solo-capability-full}
  \footnotesize
  \setlength{\tabcolsep}{4pt}
  \tablefit{%
  \begin{tabular}{@{}lccccc@{}}
    \toprule
    Model & S@7 & S@10 & RCT10 & Efficiency $E$ & MDO \\
    \midrule
    Qwen~3.8~27B
      & 0.98 [0.94, 1.00]
      & 0.98 [0.94, 1.00]
      & 5.18 [5.00, 5.42]
      & 0.97 [0.93, 1.00]
      & 0.98 [0.94, 1.00] \\
    Command~A+
      & 0.96 [0.90, 1.00]
      & 1.00 [1.00, 1.00]
      & 5.32 [5.08, 5.60]
      & 0.96 [0.92, 0.99]
      & 0.97 [0.95, 0.99] \\
    Command~R
      & 0.18 [0.08, 0.30]
      & 0.22 [0.10, 0.34]
      & 9.10 [8.54, 9.60]
      & 0.20 [0.10, 0.31]
      & 0.53 [0.48, 0.58] \\
    GPT~5.6~Sol
      & 1.00 [1.00, 1.00]
      & 1.00 [1.00, 1.00]
      & 5.00 [5.00, 5.00]
      & 1.00 [1.00, 1.00]
      & 1.00 [1.00, 1.00] \\
    \bottomrule
  \end{tabular}%
  }
\end{table}

\begin{table}[H]
  \centering
  \caption{Solo parse and retry diagnostics over recorded LLM steps.
  Cells report estimate [95\% CI]. Unparseable outputs were not observed.
  Command~R produced the single invalid move in the corpus.}
  \label{tab:solo-diagnostics}
  \footnotesize
  \setlength{\tabcolsep}{4pt}
  \begin{tabular}{@{}lccc@{}}
    \toprule
    Model & Invalid-move rate & Unparseable rate & Retry rate \\
    \midrule
    Qwen~3.8~27B
      & 0.00 [0.00, 0.00]
      & 0.00 [0.00, 0.00]
      & 0.00 [0.00, 0.00] \\
    Command~A+
      & 0.00 [0.00, 0.00]
      & 0.00 [0.00, 0.00]
      & 0.01 [0.00, 0.03] \\
    Command~R
      & 0.00 [0.00, 0.01]
      & 0.00 [0.00, 0.00]
      & 0.19 [0.15, 0.23] \\
    GPT~5.6~Sol
      & 0.00 [0.00, 0.00]
      & 0.00 [0.00, 0.00]
      & 0.00 [0.00, 0.00] \\
    \bottomrule
  \end{tabular}
\end{table}

\subsection{Initial Control Allocation}
\label{app:initial-control}

\begin{table}[H]
  \centering
  \caption{Two-way cluster-bootstrap 95\% confidence intervals for the
  first-action point estimates in Table~\ref{tab:main-allocation}, Panel~A.
  Each condition contains 40 episodes.}
  \label{tab:initial-actions-full}
  \footnotesize
  \setlength{\tabcolsep}{6pt}
  \tablefit{%
  \begin{tabular}{@{}lrccc@{}}
    \toprule
    Model & $\Delta_0$ &
    \shortstack{Self interval\\ISR} &
    \shortstack{Guide interval\\IGR} &
    \shortstack{Defer interval\\IDR} \\
    \midrule
    Qwen~3.8~27B & $-2$ & [1.00, 1.00] & [0.00, 0.00] & [0.00, 0.00] \\
                      & $0$  & [1.00, 1.00] & [0.00, 0.00] & [0.00, 0.00] \\
                      & $+2$ & [0.00, 0.40] & [0.60, 1.00] & [0.00, 0.00] \\
    \addlinespace
    Command~A+       & $-2$ & [0.83, 1.00] & [0.00, 0.18] & [0.00, 0.00] \\
                      & $0$  & [0.90, 1.00] & [0.00, 0.15] & [0.00, 0.00] \\
                      & $+2$ & [0.03, 0.50] & [0.53, 0.95] & [0.00, 0.00] \\
    \addlinespace
    Command~R        & $-2$ & [0.47, 0.90] & [0.10, 0.53] & [0.00, 0.00] \\
                      & $0$  & [0.23, 0.70] & [0.30, 0.78] & [0.00, 0.00] \\
                      & $+2$ & [0.33, 0.80] & [0.20, 0.63] & [0.00, 0.00] \\
    \addlinespace
    GPT~5.6~Sol      & $-2$ & [1.00, 1.00] & [0.00, 0.00] & [0.00, 0.00] \\
                      & $0$  & [1.00, 1.00] & [0.00, 0.00] & [0.00, 0.00] \\
                      & $+2$ & [0.00, 0.38] & [0.60, 1.00] & [0.00, 0.00] \\
    \bottomrule
  \end{tabular}%
  }
\end{table}

\begin{table}[H]
  \centering
  \caption{Two-way cluster-bootstrap 95\% confidence intervals for the
  claim-focused point estimates in Table~\ref{tab:main-allocation}, Panel~B.
  Definitions are given in Table~\ref{tab:metric-reference}.}
  \label{tab:initial-claims-full}
  \footnotesize
  \setlength{\tabcolsep}{5pt}
  \tablefit{%
  \begin{tabular}{@{}lccccc@{}}
    \toprule
    Model & ZSR & ZSB & NFSR & HFU & SSC \\
    \midrule
    Qwen~3.8~27B
      & [1.00, 1.00] & [0.50, 0.50] & [1.00, 1.00]
      & [0.58, 1.00] & [0.60, 1.00] \\
    Command~A+
      & [0.85, 1.00] & [0.35, 0.50] & [0.90, 1.00]
      & [0.50, 0.98] & [0.45, 0.95] \\
    Command~R
      & [0.22, 0.75] & [$-$0.28, 0.23] & [0.36, 0.75]
      & [0.20, 0.65] & [$-$0.29, 0.28] \\
    GPT~5.6~Sol
      & [1.00, 1.00] & [0.50, 0.50] & [1.00, 1.00]
      & [0.60, 1.00] & [0.60, 1.00] \\
    \bottomrule
  \end{tabular}%
  }
\end{table}

\subsection{Whole-Game Collaboration and Performance}
\label{app:whole-game}

\begin{table}[H]
  \centering
  \caption{Whole-game collaboration and decision quality. Action shares are
  calculated within episode and then averaged. Cells report estimate [95\% CI];
  each condition contains 40 episodes.}
  \label{tab:whole-game-full}
  \footnotesize
  \setlength{\tabcolsep}{3pt}
  \tablefit{%
  \begin{tabular}{@{}lrccccc@{}}
    \toprule
    Model & $\Delta_0$ & Self share $q^S$ & Guide share $q^G$ &
    Defer share $q^D$ & Actor aligned AAR & Direction optimal MDO \\
    \midrule
    Qwen~3.8~27B & $-2$ & 1.00 [1.00, 1.00] & 0.00 [0.00, 0.00] & 0.00 [0.00, 0.00] & 1.00 [1.00, 1.00] & 1.00 [1.00, 1.00] \\
                      & $0$  & 0.97 [0.91, 1.00] & 0.03 [0.00, 0.09] & 0.00 [0.00, 0.00] & 0.95 [0.85, 1.00] & 1.00 [1.00, 1.00] \\
                      & $+2$ & 0.03 [0.00, 0.07] & 0.97 [0.93, 1.00] & 0.00 [0.00, 0.00] & 0.97 [0.93, 1.00] & 0.97 [0.92, 1.00] \\
    \addlinespace
    Command~A+       & $-2$ & 0.98 [0.93, 1.00] & 0.02 [0.00, 0.05] & 0.01 [0.00, 0.03] & 0.98 [0.94, 1.00] & 0.92 [0.83, 0.98] \\
                      & $0$  & 0.97 [0.89, 1.00] & 0.03 [0.00, 0.11] & 0.01 [0.00, 0.03] & 0.97 [0.93, 1.00] & 0.98 [0.94, 1.00] \\
                      & $+2$ & 0.23 [0.13, 0.32] & 0.75 [0.64, 0.84] & 0.03 [0.01, 0.06] & 0.77 [0.68, 0.86] & 0.82 [0.75, 0.89] \\
    \addlinespace
    Command~R        & $-2$ & 0.39 [0.28, 0.49] & 0.61 [0.50, 0.72] & 0.01 [0.00, 0.02] & 0.51 [0.40, 0.63] & 0.39 [0.31, 0.47] \\
                      & $0$  & 0.34 [0.22, 0.49] & 0.65 [0.51, 0.76] & 0.01 [0.00, 0.03] & 0.47 [0.33, 0.64] & 0.39 [0.28, 0.50] \\
                      & $+2$ & 0.34 [0.24, 0.44] & 0.65 [0.54, 0.74] & 0.01 [0.00, 0.03] & 0.59 [0.47, 0.70] & 0.47 [0.36, 0.59] \\
    \addlinespace
    GPT~5.6~Sol      & $-2$ & 1.00 [1.00, 1.00] & 0.00 [0.00, 0.00] & 0.00 [0.00, 0.00] & 1.00 [1.00, 1.00] & 0.99 [0.95, 1.00] \\
                      & $0$  & 1.00 [1.00, 1.00] & 0.00 [0.00, 0.00] & 0.00 [0.00, 0.00] & 1.00 [1.00, 1.00] & 1.00 [1.00, 1.00] \\
                      & $+2$ & 0.05 [0.00, 0.13] & 0.95 [0.88, 1.00] & 0.00 [0.00, 0.00] & 0.95 [0.88, 1.00] & 0.93 [0.85, 1.00] \\
    \bottomrule
  \end{tabular}%
  }
\end{table}

\begin{table}[H]
  \centering
  \caption{Team outcomes by model and initial route advantage. Cells report estimate
  [95\% CI]; each condition contains 40 episodes. Lower RCT10 and higher
  efficiency $E$ are better.}
  \label{tab:team-outcomes-full}
  \footnotesize
  \setlength{\tabcolsep}{4pt}
  \tablefit{%
  \begin{tabular}{@{}lrcccc@{}}
    \toprule
    Model & $\Delta_0$ & S@7 & S@10 & RCT10 & Efficiency $E$ \\
    \midrule
    Qwen~3.8~27B & $-2$ & 1.00 [1.00, 1.00] & 1.00 [1.00, 1.00] & 5.00 [5.00, 5.00] & 1.00 [1.00, 1.00] \\
                      & $0$  & 1.00 [1.00, 1.00] & 1.00 [1.00, 1.00] & 5.20 [5.00, 5.60] & 0.97 [0.91, 1.00] \\
                      & $+2$ & 1.00 [1.00, 1.00] & 1.00 [1.00, 1.00] & 5.18 [5.00, 5.40] & 0.97 [0.93, 1.00] \\
    \addlinespace
    Command~A+       & $-2$ & 0.93 [0.80, 1.00] & 0.98 [0.85, 1.00] & 5.70 [5.25, 6.30] & 0.90 [0.79, 0.97] \\
                      & $0$  & 0.98 [0.85, 1.00] & 1.00 [1.00, 1.00] & 5.30 [5.00, 5.80] & 0.96 [0.90, 1.00] \\
                      & $+2$ & 0.70 [0.45, 0.90] & 0.88 [0.70, 1.00] & 6.78 [5.95, 7.63] & 0.71 [0.54, 0.84] \\
    \addlinespace
    Command~R        & $-2$ & 0.05 [0.00, 0.18] & 0.18 [0.00, 0.40] & 9.68 [9.10, 10.00] & 0.11 [0.00, 0.25] \\
                      & $0$  & 0.18 [0.00, 0.45] & 0.23 [0.00, 0.53] & 9.23 [8.05, 10.00] & 0.18 [0.02, 0.41] \\
                      & $+2$ & 0.25 [0.05, 0.50] & 0.45 [0.20, 0.68] & 8.93 [8.00, 9.73] & 0.31 [0.14, 0.49] \\
    \addlinespace
    GPT~5.6~Sol      & $-2$ & 1.00 [1.00, 1.00] & 1.00 [1.00, 1.00] & 5.15 [5.00, 5.60] & 0.98 [0.94, 1.00] \\
                      & $0$  & 1.00 [1.00, 1.00] & 1.00 [1.00, 1.00] & 5.00 [5.00, 5.00] & 1.00 [1.00, 1.00] \\
                      & $+2$ & 0.93 [0.75, 1.00] & 1.00 [1.00, 1.00] & 5.45 [5.05, 6.00] & 0.94 [0.87, 0.99] \\
    \bottomrule
  \end{tabular}%
  }
\end{table}

\begin{figure}[H]
  \centering
  \includegraphics[width=\linewidth]{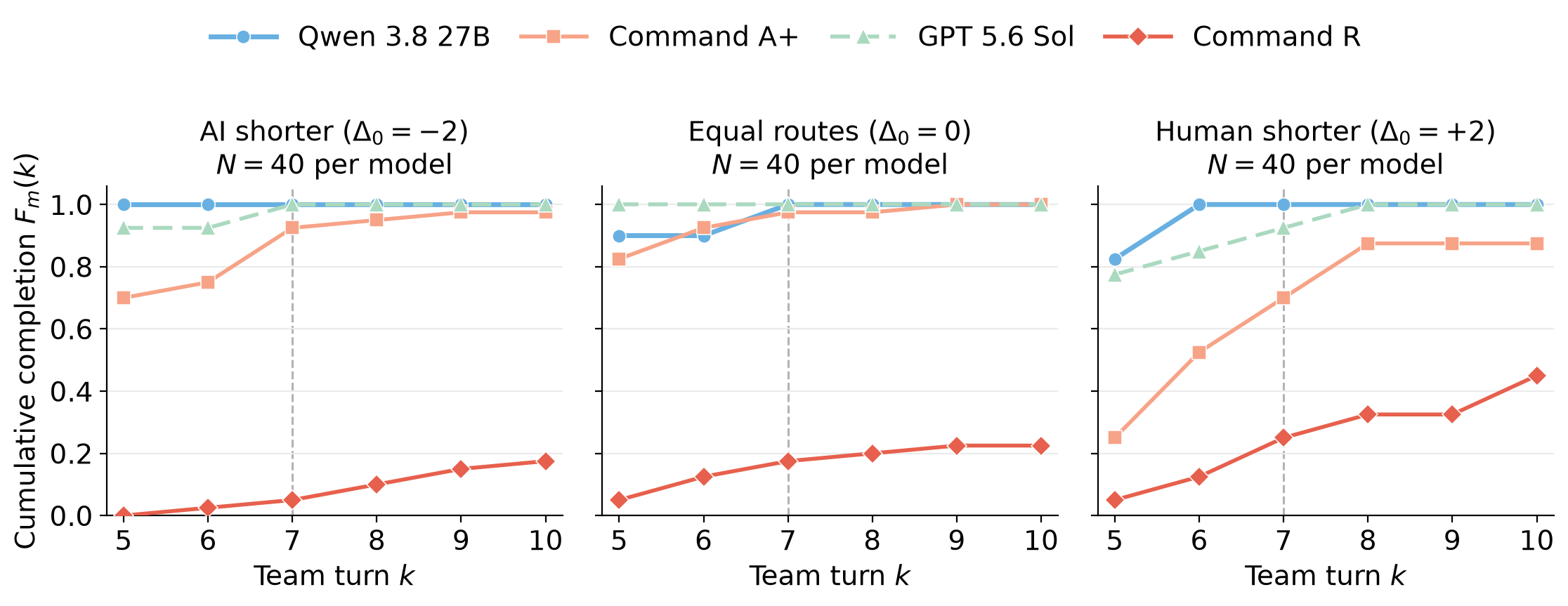}
  \caption{\textbf{Cumulative team completion by assigned route advantage.}
  Each panel plots the cumulative proportion of episodes completed by team turn
  $k$ under AI-shorter ($\Delta_0=-2$), equal-route ($\Delta_0=0$), and
  human-shorter ($\Delta_0=+2$) starting conditions. Each model--condition cell
  contains 40 episodes. The dashed vertical line marks the turn-7 checkpoint;
  episodes unfinished at turn 10 remain incomplete.}
  \label{fig:team-completion}
\end{figure}

Figure~\ref{fig:team-completion} displays cumulative team completion over the
ten-turn horizon for each assigned initial route-advantage condition.

\FloatBarrier

\newpage

\end{document}